\documentclass[format=acmsmall, nonacm=true]{acmart}

\usepackage{fullpage}
\usepackage{microtype}
\usepackage{amsmath}
\usepackage{graphicx}
\usepackage{siunitx}
\usepackage{adjustbox}

\DeclareMathOperator{\arctantwo}{arctan2}
\DeclareMathOperator{\sgn}{sgn}

\begin{document}

\acmJournal{TSAS}

\title[A Square Equal-area Map Projection]{A Square Equal-area Map Projection with\\Low Angular Distortion, Minimal Cusps, and Closed-form Solutions}
\author{Matthew~A. Petroff}
\orcid{0000-0002-4436-4215}
\affiliation{%
  \institution{Johns Hopkins University}
  \department{Department of Physics \& Astronomy}
  \streetaddress{3400 N Charles St}
  \city{Baltimore}
  \state{Maryland}
  \postcode{21218}
  \country{USA}
}
\email{petroff@jhu.edu}

\begin{abstract}
A novel square equal-area map projection is proposed. The projection combines closed-form forward and inverse solutions with relatively low angular distortion and minimal cusps, a combination of properties not manifested by any previously published square equal-area projection. Thus, the new projection has lower angular distortion than any previously published square equal-area projection with a closed-form solution. Utilizing a quincuncial arrangement, the new projection places the north pole at the center of the square and divides the south pole between its four corners; the projection can be seamlessly tiled. The existence of closed-form solutions makes the projection suitable for real-time visualization applications, both in cartography and in other areas, such as for the display of panoramic images.
\end{abstract}

\ccsdesc[500]{Human-centered computing~Geographic visualization}
\ccsdesc[100]{Computing methodologies~Image processing}

\keywords{map projection, world map, equal-area, quincuncial,\\panoramic images, sky maps}

\maketitle

\section{Introduction}

Although there is a plenitude of map projections \citep{Snyder1987}, there has been relatively little work done on square equal-area projections. As noted by \citet{Gringorten1972}, a square aspect ratio is useful for printed atlases, since it allows for a standard page to be efficiently filled with a map along with a title and descriptive caption. Equal-area projections are useful for displaying climatology and population data on the Earth, as well as for displaying cosmology data on the celestial sphere. Mappings from the sphere to the square are also useful in computer graphics applications, such as the display of panoramic images, since textures used by graphic processing units (GPUs) are often square. A quincuncial arrangement places the north pole at the center of a square and divides the south pole between its four corners, forming a quincunx pattern that resembles the ``five'' marking on a standard six-sided die. Such arrangements have previously been proposed for displaying panoramic images in an aesthetically-pleasing manner \citep{German2007} and additionally can be seamlessly tiled, which is a desirable property for texture interpolation. Equal-area projections allow for efficient use of resources, since all areas of the sphere have a constant pixel density.

The most notable existing square equal-area projection is that of \citet{Gringorten1972}, which is a quincuncial projection that was derived using differential equations, although other arrangements have also been proposed \citep{Cogley2002}. It has relatively low angular distortion and has a smooth derivative, and thus no cusps, everywhere but the equator. However, it is complicated and does not have a closed-form solution, which makes it ill-suited for use in real-time graphics display. The other significant existing square equal-area projection is what will be referred to here as the \emph{Collignon quincuncial} projection, which consists of an interrupted Collignon projection \citep{Collignon1865} for each octant of a spherical octahedron displayed in a quincuncial arrangement. It is called the \emph{equal-area zenithal orthotriangular} projection in \citet{Huang1998}, the \emph{octahedral equal area partition} in \citet{Yan2016}, and the \emph{triangular octahedral equal area} projection in \citet{McGlynn2019}. It is briefly described as a possible variant of the Collignon projection, with neither a specific name nor a figure, in \citet{Snyder1993} and is described with neither a specific name, a figure, nor a reference to the Collignon projection in \citet{Maurer1935} (translated to English in \citet{Warntz1968}); it is also described for a single hemisphere in \citet{Rosca2011} and \citet{Holhos2014}. This projection has the advantage of being mathematically simple and has a closed-form solution, for both forward and inverse mappings. However, it has larger angular distortions than the Gringorten projection and has significant visible cusps along the edges of each projected octahedron face. Both of these projections, as with all quincuncial projections, can be thought of as polyhedral map projections for the octahedron with an additional linear rescaling along the axis perpendicular to the equator. Square equal-area projections with non-quincuncial arrangments, e.g., square aspect ratio cylindrical equal-area projections \citep{Close1947, Tobler1986}, will not be considered, since they have fewer interruptions and thus much higher angular distortion.

Therefore, to improve upon existing square equal-area projections, a new projection must combine the low angular distortion of the Gringorten projection and its reduced occurrence of significant visible cusps with the Collignon quincuncial projection's advantage of having closed-form solutions. These are the properties of the new projection presented in this manuscript. The vertex-oriented great circle projection ``slice-and-dice'' technique of \citet{vanLeeuwen2006} is generalized for the octahedron such that the subdivision of each face can be optimized to minimize angular distortion when each octant of a spherical octahedron is mapped to an isosceles right triangle instead of to an equilateral triangle, while still maintaining the equal-area property of the mapping; these isosceles right triangles are then displayed in a quincuncial arrangement. An octahedron is used, since it is the only polyhedron suitable for mapping onto a square in a quincuncial arrangement. A square equal-area projection without discontinuities can also be constructed from a tetrahedron, but such a projection would have increased angular distortion and would have a different arrangement; the other Platonic solids cannot be mapped onto the square without discontinuities. The new projection is described in detail in Section~\ref{sec:projection}, the properties of the projection are discussed in Section~\ref{sec:discussion}, and the article concludes in Section~\ref{sec:conclusions}.

\section{Projection}
\label{sec:projection}

\begin{figure}
\centering
\includegraphics[width=\textwidth]{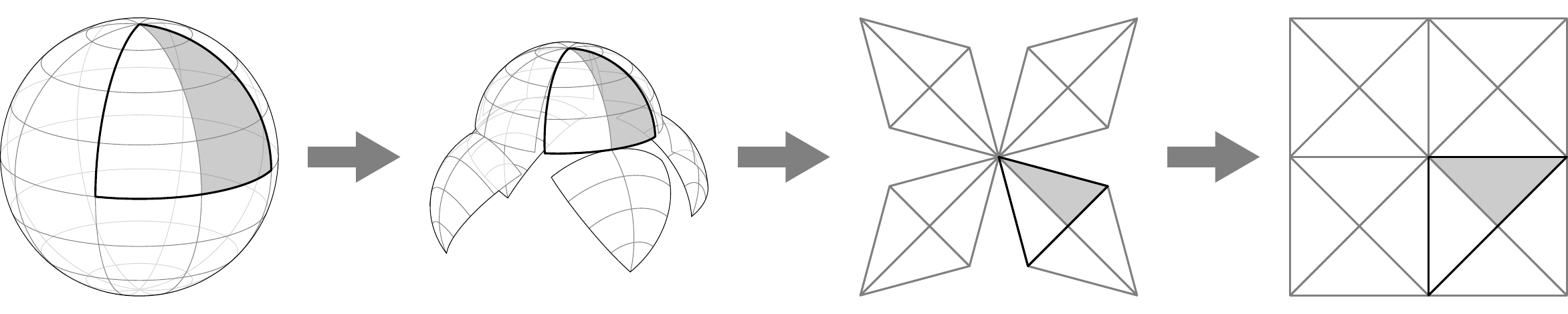}
\Description{A progression of line drawings showing a sphere being unwrapped and squished into a square, with an octant and a hexadecant highlighted.}
\caption{Outline of projection steps. Starting with an octant and a hexadecant highlighted on the sphere, the sphere is unwrapped and split by octant. Each octant is then flattened onto an equilateral triangle such that the octant's left hexadecant is a mirror of the octant's right hexadecant and such that area ratios are preserved; this step comprises the majority of the projection's details and involves further subdividing each hexadecant into three sub-triangles. Finally, the equilateral triangles are squished and rearranged into a square, in a quincuncial arrangement.}
\label{fig:outline}
\end{figure}

Before presenting the new projection, an outline of its derivation will be given. The projection is based on a procedure of mapping each octant of a spherical octahedron onto a Euclidean equilateral triangle using the following steps:
\begin{enumerate}
\item Each octant, both spherical and Euclidean, is subdivided into two symmetric right triangles, which are each a \emph{hexadecant}, one-sixteenth of the sphere.
\item Each hexadecant is further subdivided into three sub-triangles, which are sized such that the area ratios between each Euclidean sub-triangle and the Euclidean hexadecant match the area ratios between each spherical sub-triangle and the spherical hexadecant.
\item Each of the spherical sub-triangles is mapped onto its corresponding Euclidean sub-triangle in an equal-area manner.
\item The Euclidean hexadecants are then squished and rearranged into a square, in a quincuncial arrangement.
\end{enumerate}
An overview of these steps is shown in Figure \ref{fig:outline}.

\begin{figure}
\centering
\includegraphics[width=2in]{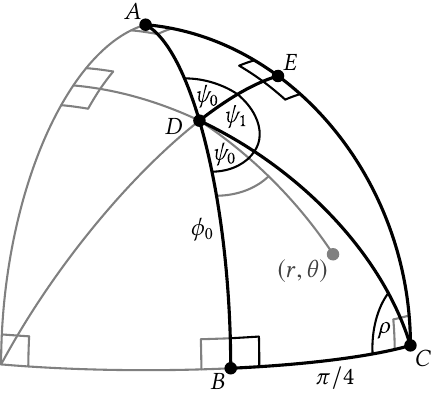}\hspace*{3em}\includegraphics[width=2in]{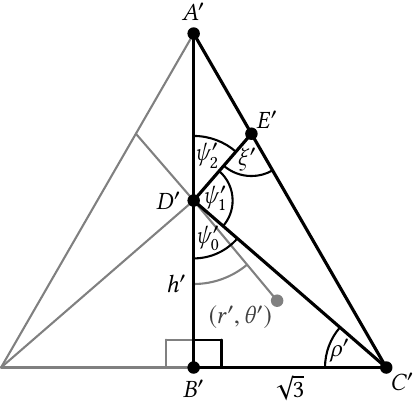}
\Description{Line drawings of spherical and Euclidean octants with angles, distances, and vertices labeled.}
\caption{Octant mapping. The octant of the spherical octahedron is shown on the left, and the corresponding equilateral Euclidean triangle is shown on the right. Angles, distances, and vertices used in the text are labeled. The right halves are highlighted, since the projection is mirrored about the centerline. The scaling of the Euclidean triangle is somewhat arbitrary but was chosen to allow for additional simplification of some of the equations in the projection derivation.}
\label{fig:facet}
\end{figure}

Now, the derivation can proceed. Here, we wish to create a mapping from latitude, $\phi\in[-\pi/2,\pi/2]$, and longitude, $\lambda \in[0,2\pi)$, to Cartesian coordinates, $x,y\in[-1,1]$. As a quincuncial arrangement is to be used, the map coordinate system will be oriented such that the north pole is at the origin and the south pole is split between the four corners. The prime meridian runs from the origin to the bottom center, ($0$,~$1$), where it intersects the equator; the meridian is then interrupted and runs along the bottom edge of the map in both directions to the bottom two corners, ($-1$,~$-1$) and ($1$,~$-1$). The \SI{\pm 90}{\degree} and \SI{180}{\degree} meridians will also be interrupted below the equator. The creation of an equal-area mapping based on an octahedron can be decomposed into the mapping of one octant of a spherical octahedron onto an equilateral triangle. Furthermore, the octant can be split into two symmetric right triangles, so only a hexadecant has to be considered. This right triangle can be further decomposed into three smaller triangles that share a vertex along the edge by which the right triangle is mirrored. This decomposition is shown in Figure~\ref{fig:facet}, which serves as a reference for angle, distance, and vertex labels used later in this derivation. In \citet{vanLeeuwen2006}, this vertex was chosen to be at the center of the spherical octahedron octant, which makes the smaller Euclidean triangles right triangles that are identical up to a mirroring about the hypotenuse. While this is optimal when the spherical octant is mapped to an equilateral triangle, it leads to increased average angular distortion when this equilateral triangle is scaled to become a right isosceles triangle, as is necessary to form a square map. Thus, the more general case is considered here, where the vertex position is allowed to move along the octant's mirroring axis.

In the more general case, the smaller triangles differ in area and are not all right triangles when mapped to the plane. For the map to remain equal-area, the area ratios of the smaller triangles to the hexadecant must be constant between the spherical triangles and the Euclidean triangles.
To set this area ratio, some angles of the sub-triangles first need to be defined. For the spherical sub-triangles, these angles are
\begin{equation}
\psi_0 = \arcsin\left( \frac{1}{\sqrt{2 - \cos^2\phi_0}} \right)
\end{equation}
\begin{equation}
\psi_1 = \pi - 2\psi_0
\end{equation}
\begin{equation}
\rho = \arcsin\left(\frac{2\sin\phi_0}{\sqrt{3-\cos(2\phi_0)}}\right),
\end{equation}
where $\phi_0$ is the latitude of the dividing point on the spherical triangle.
For the corresponding Euclidean sub-triangles, these angles are
\begin{equation}
\label{eq:psi0prime}
\psi_0' = \arctan\left(\frac{\sqrt{3}}{h'}\right)
\end{equation}
\begin{equation}
\label{eq:psi1prime}
\psi_1' = \frac{7\pi}{6} - \psi_0' - \xi'
\end{equation}
\begin{equation}
\label{eq:psi2prime}
\psi_2' = \xi' - \frac{\pi}{6}
\end{equation}
\begin{equation}
\rho' = \arctan\left(\frac{h'}{\sqrt{3}}\right),
\end{equation}
where $h'$ is the distance on the Euclidean triangle corresponding to $\phi_0$ and $\xi'$ is an angle that will be discussed later in the derivation.
Setting the area ratio is then accomplished by setting the height of triangle $B'C'D'$ such that the area ratio between triangle $B'C'D'$ and triangle $A'B'C'$ matches the area ratio between spherical triangle $BCD$ and spherical triangle $ABC$.
This produces the relation
\begin{equation}
\frac{h'\sqrt{3}/2}{3\sqrt{3}/2} = \frac{\psi_0 + \rho - \pi/2}{\pi/4},
\end{equation}
which results in
\begin{equation}
\label{eq:hprime}
h' = \frac{12}{\pi}\left( \psi_0 + \rho - \frac{\pi}{2} \right)
\end{equation}
when solved for $h'$. Next, the angle $\xi'$ is adjusted such that the area ratio between triangle $C'D'E'$ and triangle $A'B'C'$ matches the area ratio between spherical triangle $CDE$ and spherical triangle $ABC$. This produces the relation
\begin{equation}
\frac{\left(\!\sqrt{h'^2+3}\right)^2\sin\left(\pi / 3 - \rho'\right)\sin\left[(\pi / 3 - \rho') + \xi'\right] / (2\sin\xi')}{3\sqrt{3}/2} = \frac{\pi - 2\psi_0 - \rho}{\pi/4},
\end{equation}
which results in
\begin{equation}
\label{eq:xiprime}
\xi'=\arctan\left( \frac{\pi(h'-3)^2}{\sqrt{3}\left[ \pi\left(h'^2-2h'+45\right) - 96\psi_0 - 48\rho \right]} \right)
\end{equation}
when solved for $\xi'$. Since the area ratios of two of the three smaller triangles are now preserved, the area ratio of the third smaller triangle is also preserved. With the relative areas of the three smaller triangles preserved, an equal-area mapping of each of the smaller triangles will result in a mapping that is equal-area over the whole sphere.

\begin{figure}
\centering
\includegraphics[width=2in]{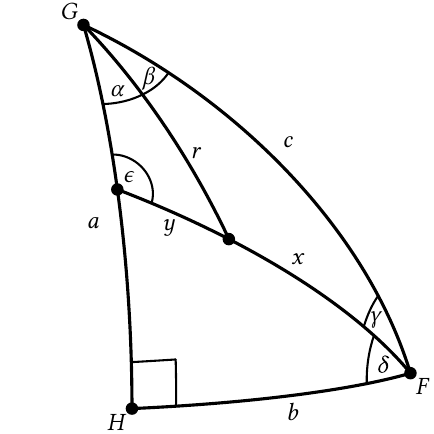}\hspace*{3em}\includegraphics[width=2in]{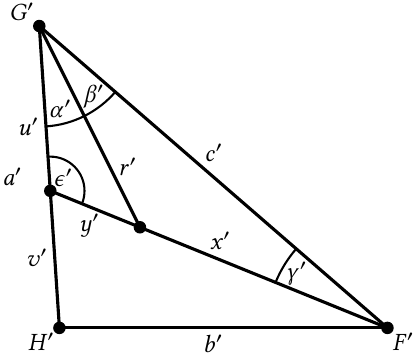}
\Description{Line drawings of spherical and Euclidean sub-triangles with angles, distances, and vertices labeled.}
\caption{Sub-triangle mapping. The spherical triangle is shown on the left, while the Euclidean triangle is shown on the right. Angles, distances, and vertices used in the text are labeled. Triangle $FGH$ ($F'G'H'$) in this figure corresponds to triangles $BCD$ ($B'C'D'$), $CDE$ ($C'D'E'$), and $ADE$ ($A'D'E'$) in Figure~\ref{fig:facet}, with side lengths and angles that vary depending on the target triangle.}
\label{fig:triangle}
\end{figure}

Next, the mapping within the sub-triangles needs to be defined. Figure~\ref{fig:triangle} serves as a reference for geometry labels for the sub-triangles; the letters used to label the sub-triangle vertices in the figure are also used to refer to the corresponding interior angles. Here, we will restrict ourselves to positive latitudes by setting $\phi_c = |\phi|$ and center longitudes around the center of each octant with $\lambda_c = \lambda - \frac{\pi}{4}$. The longitude, $\lambda_0$, of the dividing point of each octant is
\begin{equation}
q = \left\lfloor \frac{2}{\pi}\lambda \right\rfloor
\end{equation}
\begin{equation}
\lambda_0 = \frac{\pi}{2}q,
\end{equation}
where, $\left\lfloor \;\right\rfloor$ denotes the floor function. We then define each point by an angle around and a distance from the dividing point with
\begin{equation}
\theta = \left|\arctantwo\left[ \cos\phi_c \sin(\lambda_c - \lambda_0), \sin\phi_0 \cos\phi_c \cos(\lambda_c - \lambda_0) - \cos\phi_0 \sin\phi_c \right]\right|
\end{equation}
\begin{equation}
r = \arccos\left[ \sin\phi_0 \sin\phi_c + \cos\phi_0 \cos\phi_c \cos(\lambda_c - \lambda_0) \right],
\end{equation}
where $\theta$ is the counter-clockwise angle around the dividing point starting at zero at the equator and $r$ is the distance from the dividing point. We then define the angle, $\beta$, between the ray to the target point and the hypotenuse of the sub-triangle as
\begin{equation}
\beta = \begin{cases}
\psi_0 - \theta & \theta \leq \psi_0 \\
\theta - \psi_0 & \psi_0 < \theta \leq \psi_0 + \psi_1 \\
\pi - \theta & \psi_0 + \psi_1 < \theta.
\end{cases}
\end{equation}
The hypotenuse of the sub-triangle, $c$, is
\begin{equation}
\label{eq:c}
c = \begin{cases}
\arccos\left(\cos\phi_0/\sqrt{2}\right) & \theta \leq \psi_0 + \psi_1 \\
\pi/2 - \phi_0 & \theta > \psi_0 + \psi_1.
\end{cases}
\end{equation}
The angle of the spherical sub-triangle at the vertex at the dividing point is
\begin{equation}
\label{eq:G}
G = \begin{cases}
\psi_0 & \theta \leq \psi_0 \\
\psi_1 & \psi_0 < \theta \leq \psi_0 + \psi_1 \\
\psi_0 & \psi_0 + \psi_1 < \theta,
\end{cases}
\end{equation}
and
\begin{equation}
\label{eq:Gprime}
G' = \begin{cases}
\psi_0' & \theta \leq \psi_0 \\
\psi_1' & \psi_0 < \theta \leq \psi_0 + \psi_1 \\
\psi_2' & \theta > \psi_0 + \psi_1
\end{cases}
\end{equation}
is the corresponding angle on the Euclidean sub-triangle.
The angle of the spherical sub-triangle at the corner of the octant is
\begin{equation}
\label{eq:F}
F = \begin{cases}
\rho & \theta \leq \psi_0 \\
\pi/2 - \rho & \psi_0 < \theta \leq \psi_0 + \psi_1 \\
\pi/4 & \theta > \psi_0 + \psi_1.
\end{cases}
\end{equation}
The length of the side of the Euclidean sub-triangle extending from the mid-point of the octant side to the dividing point is
\begin{equation}
\label{eq:aprime}
a' = \begin{cases}
h' & \theta \leq \psi_0 \\
\sqrt{h'^2 + 3}\sin\left( \pi/3 - \rho' \right)/\sin\xi' & \theta > \psi_0,
\end{cases}
\end{equation}
and
\begin{equation}
\label{eq:cprime}
c' = \begin{cases}
\sqrt{h'^2+3} & \theta \leq \psi_0 + \psi_1 \\
3 - h' & \theta > \psi_0 + \psi_1
\end{cases}
\end{equation}
is the hypotenuse length of the same Euclidean sub-triangle. The distance from the octant corner to the target point on the spherical sub-triangle is
\begin{equation}
x = \arccos(\cos r \cos c + \sin r \sin c \cos\beta),
\end{equation}
and the angle between that ray and the hypotenuse of the spherical sub-triangle is
\begin{equation}
\gamma = \arcsin\left(\frac{\sin\beta\sin r}{\sin x}\right).
\end{equation}
If that ray is extended to meet the side of the spherical triangle that extends from the mid-point of the octant side to the dividing point, the angle between that ray and the ray extending from the intersection point to the dividing point is
\begin{equation}
\epsilon = \arccos\left( \sin G \sin\gamma \cos c - \cos G \cos\gamma \right).
\end{equation}

With these angles and lengths defined, we can proceed with the ``slice-and-dice'' approach. First, we divide the side of the Euclidean triangle that extends from the mid-point of the octant side to the dividing point into lengths $u'$ and $v'$, and we divide the line that extends from the octant corner to this side into lengths $x'$ and $y'$. We then define the ratio of the areas of the two slices of the Euclidean sub-triangle to match that of the spherical sub-triangle using
\begin{equation}
\frac{u'}{(u'+v')} = \frac{\gamma + G + \epsilon - \pi}{F + G - \frac{\pi}{2}}.
\end{equation}
Next, we ``dice'' these slices by defining the separation point on the slicing line to also maintain the area ratios using
\begin{equation}
\begin{split}
\cos(x + y) &= \cos\left[\arcsin\left(\frac{\sin G\sin c}{\sin\epsilon}\right)\right] \\
&= \sqrt{1 - \left(\frac{\sin G\sin c}{\sin\epsilon}\right)^2}
\end{split}
\end{equation}
\begin{equation}
\label{eq:dice}
\frac{x'}{(x'+y')} = \sqrt{\frac{1 - \cos x}{1 - \cos(x + y)}}.
\end{equation}
Solving for a few remaining lengths and angles,
\begin{equation}
u' = (a')\left(\frac{u'}{(u'+v')}\right)
\end{equation}
\begin{equation}
(x' + y') = \sqrt{u'^2 + c'^2 - 2u'c'\cos G'}
\end{equation}
\begin{equation}
\begin{split}
\cos\gamma' &= \cos\left[\arcsin\left(\frac{u'\sin G'}{(x' + y')}\right)\right] \\
&= \sqrt{1 - \left(\frac{u'\sin G'}{(x' + y')}\right)^2}
\end{split}
\end{equation}
\begin{equation}
x' = (x' + y')\left(\frac{x'}{(x' + y')}\right)
\end{equation}
\begin{equation}
y' = (x' + y') - x',
\end{equation}
we can define the target point on the Euclidean triangle in polar coordinates with respect to the dividing point with
\begin{equation}
r' = \sqrt{x'^2 + c'^2 - 2x'c'\cos\gamma'}
\end{equation}
\begin{equation}
\alpha' = \arccos\left(\frac{y'^2 - u'^2 - r'^2}{-2u'r'}\right).
\end{equation}
We then convert this angle from coordinates relative to the sub-triangle to coordinates relative to the octant,
\begin{equation}
\theta' = \begin{cases}
\alpha' & \theta \leq \psi_0 \\
\frac{7\pi}{6} - \xi' - \alpha' & \psi_0 < \theta \leq \psi_0 + \psi_1 \\
\frac{7\pi}{6} - \xi' + \alpha' & \psi_0 + \psi_1 < \theta.
\end{cases}
\end{equation}
Then, we convert to Cartesian coordinates with
\begin{align}
x_c &= \sgn(\lambda_c - \lambda_0)r'\sin\theta' \\
y_c &= h' - r'\cos\theta'
\end{align}
and separate northern and southern hemispheres with
\begin{equation}
y_h = y_c \sgn\phi - 3.
\end{equation}
Finally, we squish the equilateral triangle into a right isosceles triangle and arrange the octants into a square with
\begin{equation}
\zeta = \frac{\pi}{4} + \frac{\pi}{2}q
\end{equation}
\begin{equation}
\begin{pmatrix}
x_m \\ y_m
\end{pmatrix} = \begin{pmatrix}
x_c \cos\zeta - y_h\sin\zeta/\sqrt{3} \\
x_c \sin\zeta + y_h\cos\zeta/\sqrt{3}
\end{pmatrix}\frac{\sqrt{3}}{3\sqrt{2}}.
\end{equation}
As the ratio of areas is an affine invariant, this operation maintains the equal-area property of the projection.

The maximum angular distortion at a given point, $\omega$, is calculated using Tissot's indicatrix \citep{Tissot1881, Snyder1987}. This is defined using
\begin{align}
h &= \sqrt{\left(\frac{\partial x}{\partial\phi}\right)^2 + \left(\frac{\partial y}{\partial\phi}\right)^2} \\
k &= \sqrt{\left(\frac{\partial x}{\partial\lambda}\right)^2 + \left(\frac{\partial y}{\partial\lambda}\right)^2} \\
\sin\eta' &= \frac{1}{hk\cos\phi}\left[\left(\frac{\partial y}{\partial\phi}\right)\left(\frac{\partial x}{\partial\lambda}\right) - \left(\frac{\partial x}{\partial\phi}\right)\left(\frac{\partial y}{\partial\lambda}\right)\right] \\
\label{eq:distort}
\omega &= 2\arcsin\left(\sqrt{\frac{h^2 + k^2 - 2hk\sin\eta'}{h^2 + k^2 + 2hk\sin\eta'}}\right),
\end{align}
where $h$ and $k$ are the scale factors along the meridian and parallel, respectively, and $\eta'$ is the angle at which a given meridian and parallel intersect. In order to reduce this distortion, numerical optimization was used to find the value of $\phi_0$ that minimized the average of $\omega$ across the entire map; the exact procedure used to calculate this average angular distortion is discussed in the next section. This result was very nearly $3\pi/8$, so $\phi_0$ was set to this exact value. Tissot's indicatrix can also be used to calculate the area scale factor, $s$, using
\begin{equation}
\label{eq:area}
s = hk\sin\eta'.
\end{equation}
As the projection is equal-area, this value is constant throughout the map. Using Equation~(\ref{eq:area}) results in a area scale factor of $s\approx 0.318$. Alternatively, the area scale factor can be computed via the ratio of the unitless area of the map to the area of the unit sphere in steradians, $4\pi$. Since the projection maps to $x,y\in[-1,1]$, this gives an area scale factor of $2^2/(4\pi)=1/\pi\approx0.318$. As expected, this is consistent with the value calculated using Tissot's indicatrix.

\begin{figure}
\centering
\includegraphics[width=\textwidth]{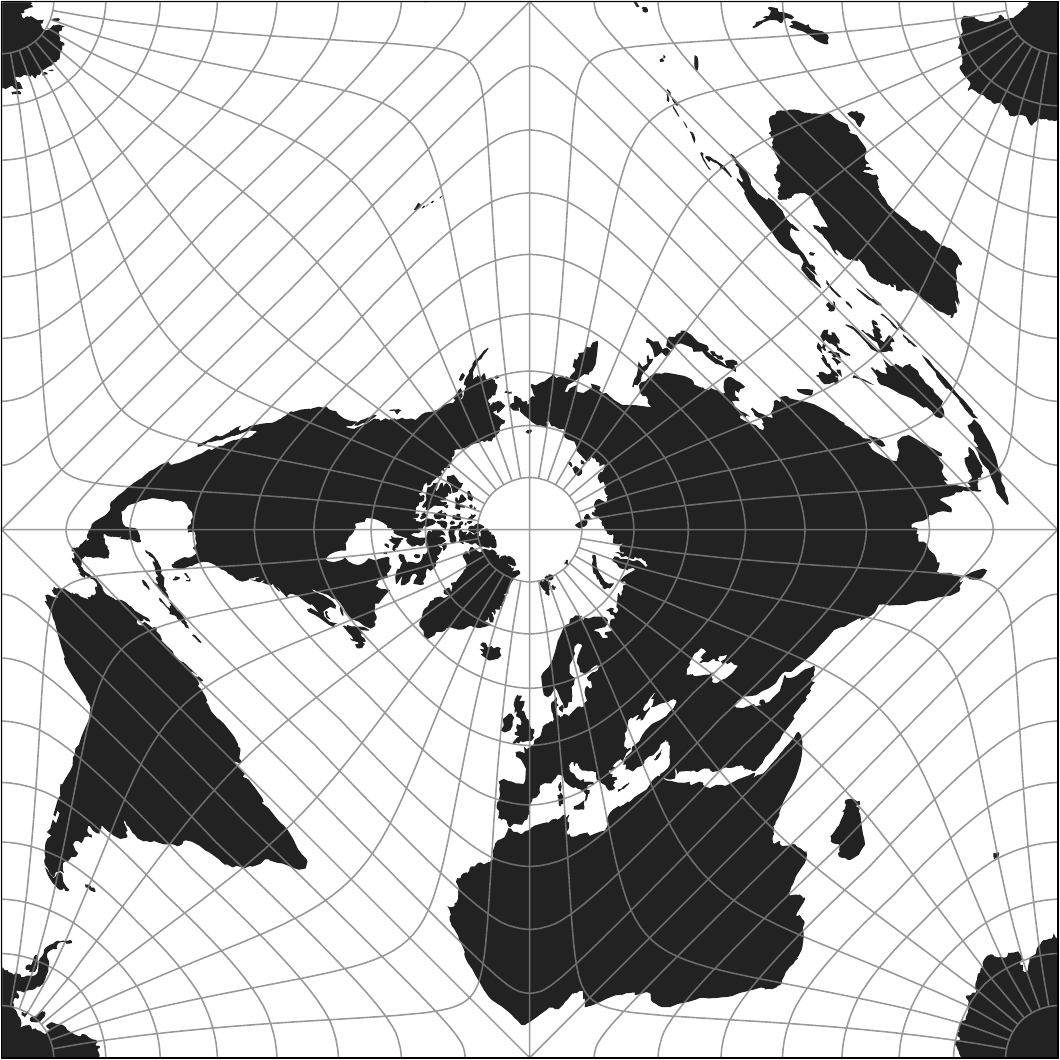}
\Description{A quincuncial world map with land masses drawn in black and a graticule overlaid in gray.}
\caption{A world map constructed using the new projection presented in this work. A \SI{10}{\degree} graticule is overlaid on the map. Note the lack of significant visible cusps.}
\label{fig:map}
\end{figure}

A world map constructed using the projection is shown in Figure~\ref{fig:map}. Although the above equations are not directly invertible, the vertex-oriented great circle projection technique does lend itself to a closed-form inverse mapping, and a closed-form inverse mapping of the projection just presented is given in Appendix~\ref{app:inverse}.

\section{Discussion}
\label{sec:discussion}

As presented in \citet{vanLeeuwen2006}, the vertex-oriented great circle projection technique is symmetric around the center of each triangle it projects. Although this is optimal for mapping equilateral triangles, it inflates the angular distortion for isosceles triangles. The generalization of said technique in this work, while maintaining its equal-area property, allows it to be optimized for isosceles triangles, such as the isosceles right triangles that subdivide a quincuncial projection. In addition to its use in the quincuncial projection presented in this work, the generalized technique can also be used to produce optimized equal-area star projections, which can be constructed by mapping to isosceles triangles.

\begin{figure}
\centering
\begin{tabular}{c@{\hskip 0.75em}c@{\hskip 0.75em}c@{\hskip 0.1em}c}
\small{\textsf{Collignon quincuncial}} & \small{\textsf{Gringorten}} & \small{\textsf{This work}} \\
\includegraphics[height=33mm]{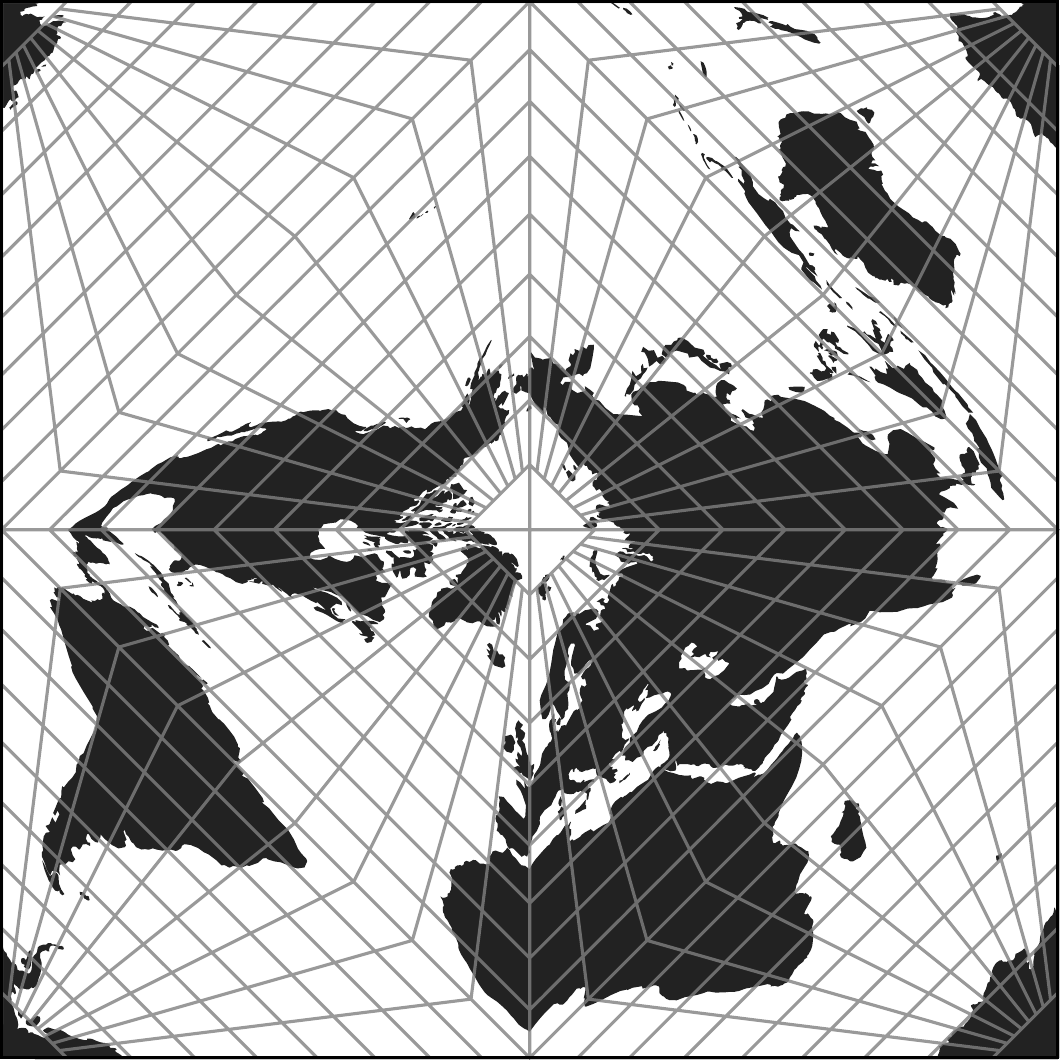} & \includegraphics[height=33mm]{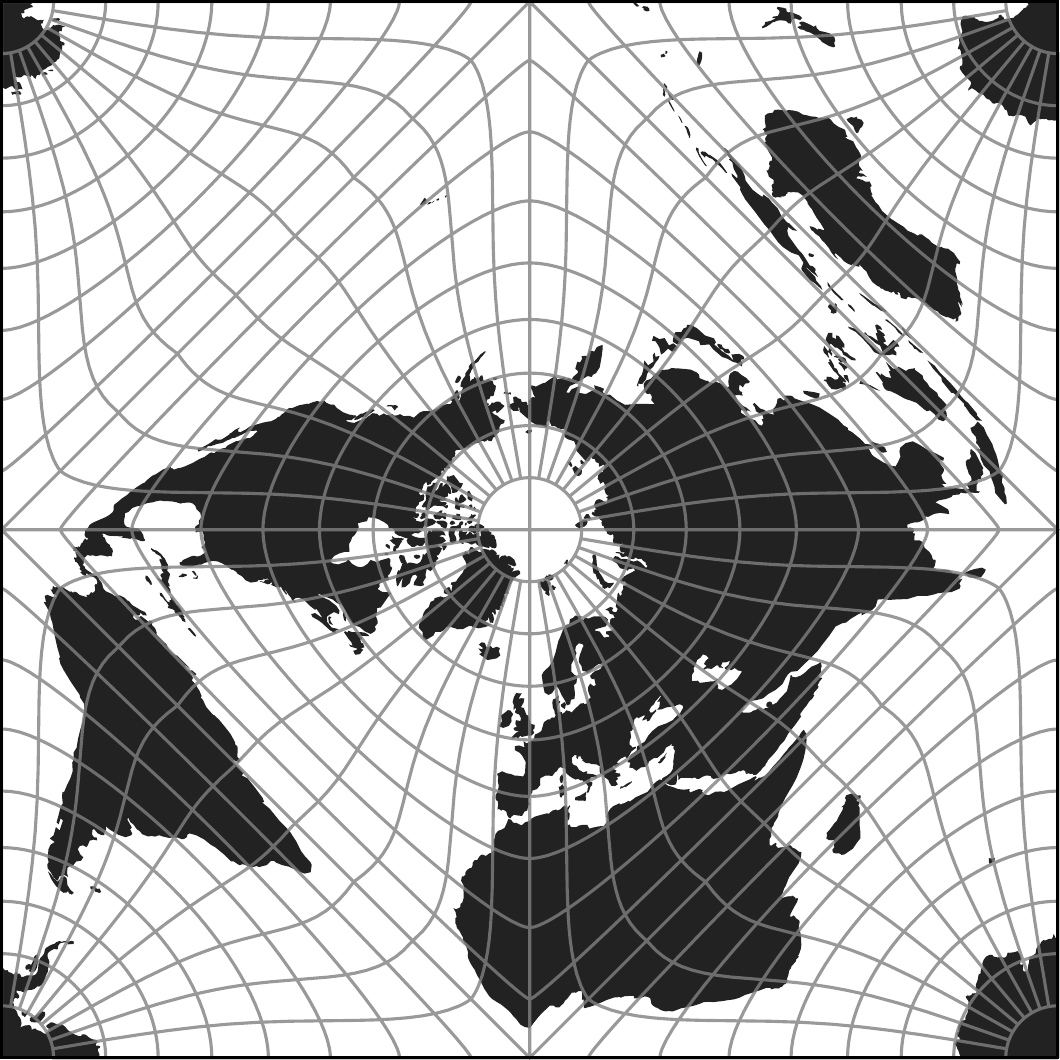} & \includegraphics[height=33mm]{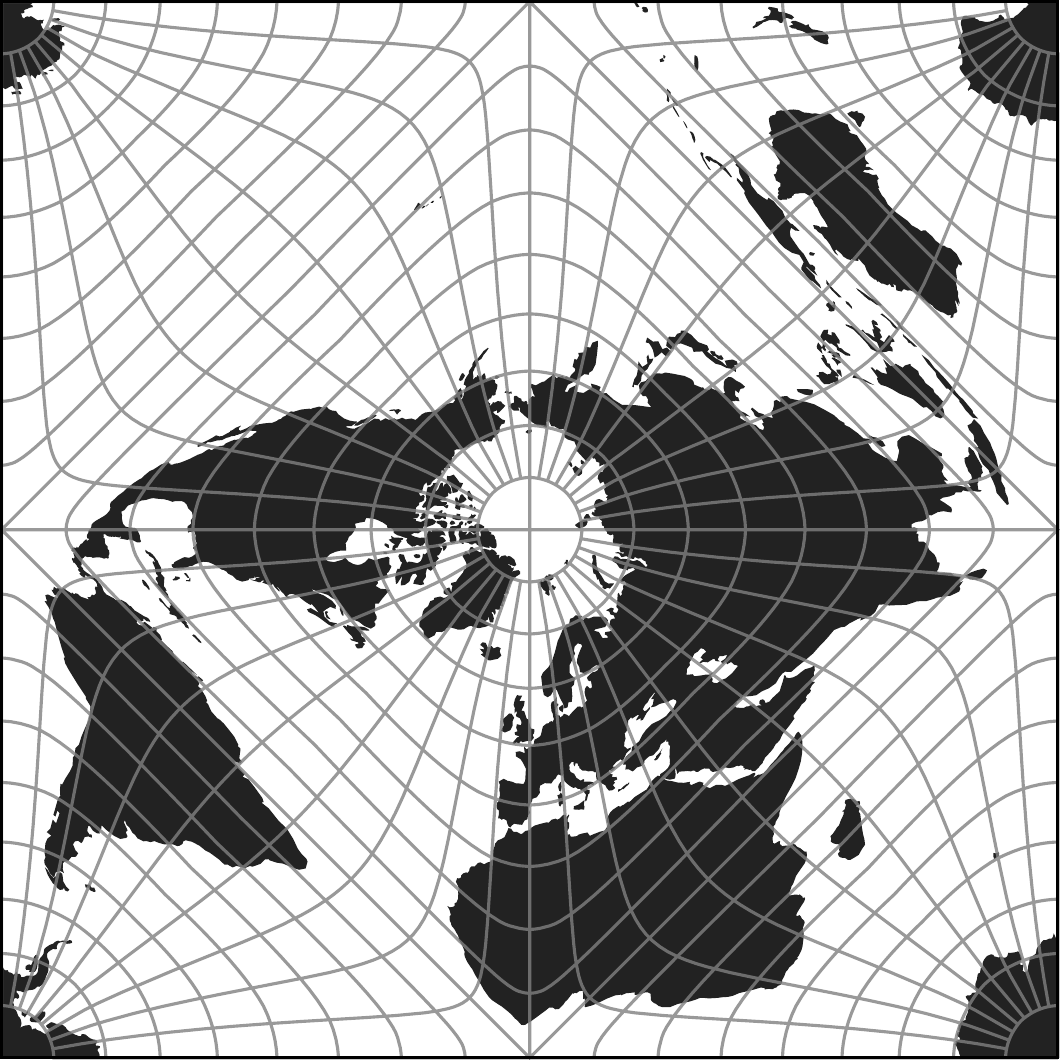} \\
\adjustbox{width=33mm, clip=true,trim=13 8 18 28}{\includegraphics[height=10mm, angle=45]{includes/collignon-quincuncial.pdf}} & \adjustbox{width=33mm, clip=true,trim=13 8 18 28}{\includegraphics[height=10mm, angle=45]{includes/gringorten.pdf}} & \adjustbox{width=33mm, clip=true,trim=13 8 18 28}{\includegraphics[height=10mm, angle=45]{includes/new-projection.pdf}} \\
\includegraphics[height=33mm]{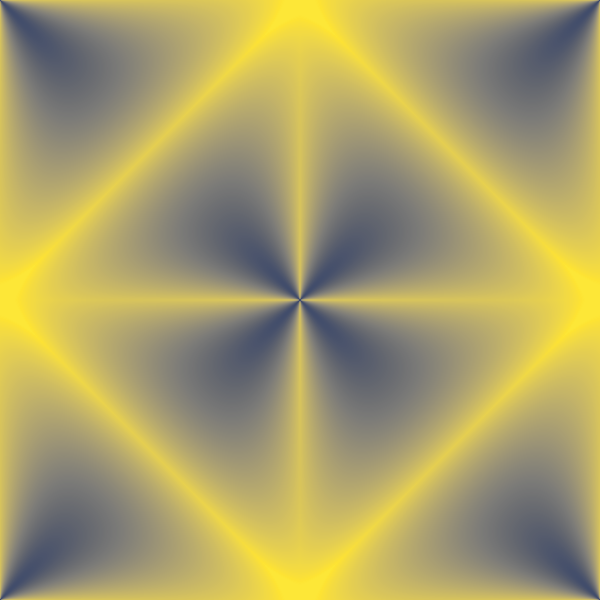} & \includegraphics[height=33mm]{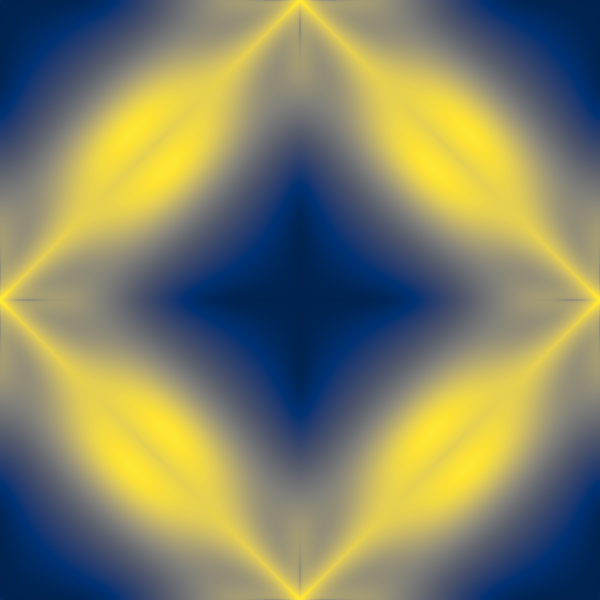} & \includegraphics[height=33mm]{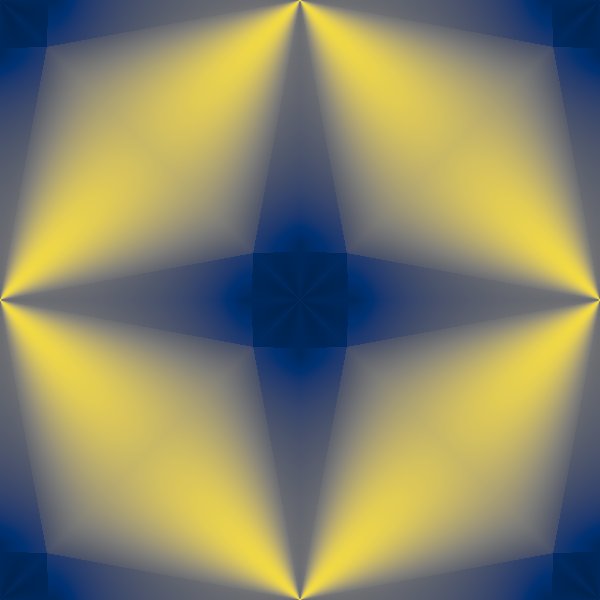} & \includegraphics[height=33mm]{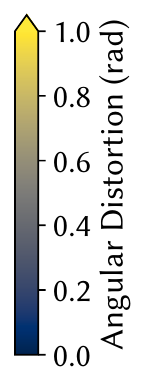}
\end{tabular}
\Description[Three quincuncial world maps are shown with corresponding angular distortion maps.]{Three quincuncial world maps are shown with corresponding angular distortion maps. The Collignon quincuncial projection angular distortion map shows higher distortions than the other two angular distortion maps. The Gringorten projection angular distortion map shows distortions more concentrated near the equator.}
\caption{Map projection comparison. The new projection presented in this work is compared to the existing Collignon quincuncial and Gringorten projections. The top row shows a world map; the middle row shows an excerpt of the world map near the equator displaying South America, which compares the degree to which cusps are visible; and the bottom row shows the angular distortion ($\omega$), as calculated using Tissot's indicatrix. The new projection presented in this work has considerably smaller cusps at the equator than the other two projections, and its angular distortion is more spread out than the Gringorten projection.}
\label{fig:maps}
\end{figure}

A comparison of the new projection presented in this work with the Collignon quincuncial and Gringorten projections is shown in Figure~\ref{fig:maps}, which includes a world map and an angular distortion map, as calculated using Tissot's indicatrix [Equation~(\ref{eq:distort})], for each projection.\footnote{The distortion was calculated using automatic (analytic) differentiation for the Collignon quincuncial projection and the new projection presented in this work but was calculated using a finite-difference method for the Gringorten projection, since it needs to be implemented using iterative methods.} The forward mapping equations used for the Collignon quincuncial projection are presented in Appendix~\ref{app:collignon}; for the Gringorten projection equations, the reader is referred to \citep{Gringorten1972}. As can be seen in this comparison, the new projection contains no significant visible cusps, which improves upon the Gringorten projection's cusp at the equator and the Collignon quincuncial projection's significant equatorial and meridional cusps. All three projections can be seamlessly tiled, with adjacent tiles rotated by \SI{180}{\degree} (\emph{p2} wallpaper group). To derive distortion statistics, the distortion was calculated at \num{10000} points placed using a Fibonacci lattice \citep{Baselga2018}. The average, standard deviation, and maximum angular distortion values are given in Table~\ref{tab:distort} for the three projections. Both the new projection and the Gringorten projection have significantly lower average angular distortion than the Collignon quincuncial projection. The angular distortion of the new projection and the Gringorten projection are comparable, with the new projection having lower maximum distortion but the Gringorten projection having slightly lower average distortion; the new projection's angular distortion is more spread out than the Gringorten projection, which concentrates angular distortion near the equator. As all three projections are equal-area, comparison of area distortions is moot.

\begin{table}
\caption{Angular Distortion Summary}
\label{tab:distort}
\begin{tabular}{rSSS}
\toprule
Projection & {Average (rad)} & {Standard deviation (rad)} & {Max (rad)} \\
\midrule
Collignon quincuncial & 0.68 & 0.18 & 1.05 \\
Gringorten & 0.52 & 0.31 & 1.05 \\
This work & 0.54 & 0.27 & 0.95 \\
\bottomrule
\end{tabular}
\end{table}

Excluding constants, which can be pre-computed, and eliminating trivial identities, e.g., $z=\cos(\arccos z)$, the given forward mapping equations of the new projection require 12 trigonometric function evaluations and 6 inverse trigonometric function evaluations; four of the trigonometric function evaluations can be further eliminated using identities for the composition of trigonometric and inverse trigonometric functions, bringing the total number of required trigonometric function evaluations down to eight. While more computationally complex than the Collignon quincuncial projection, which requires only a single trigonometric function evaluation in its forward mapping, the new projection still has closed-form forward and inverse solutions, which allow it to be implemented as a GPU shader for real-time visualizations, unlike the iterative methods required for the Gringorten projection.

\begin{figure}
\centering
\includegraphics[width=2.1in]{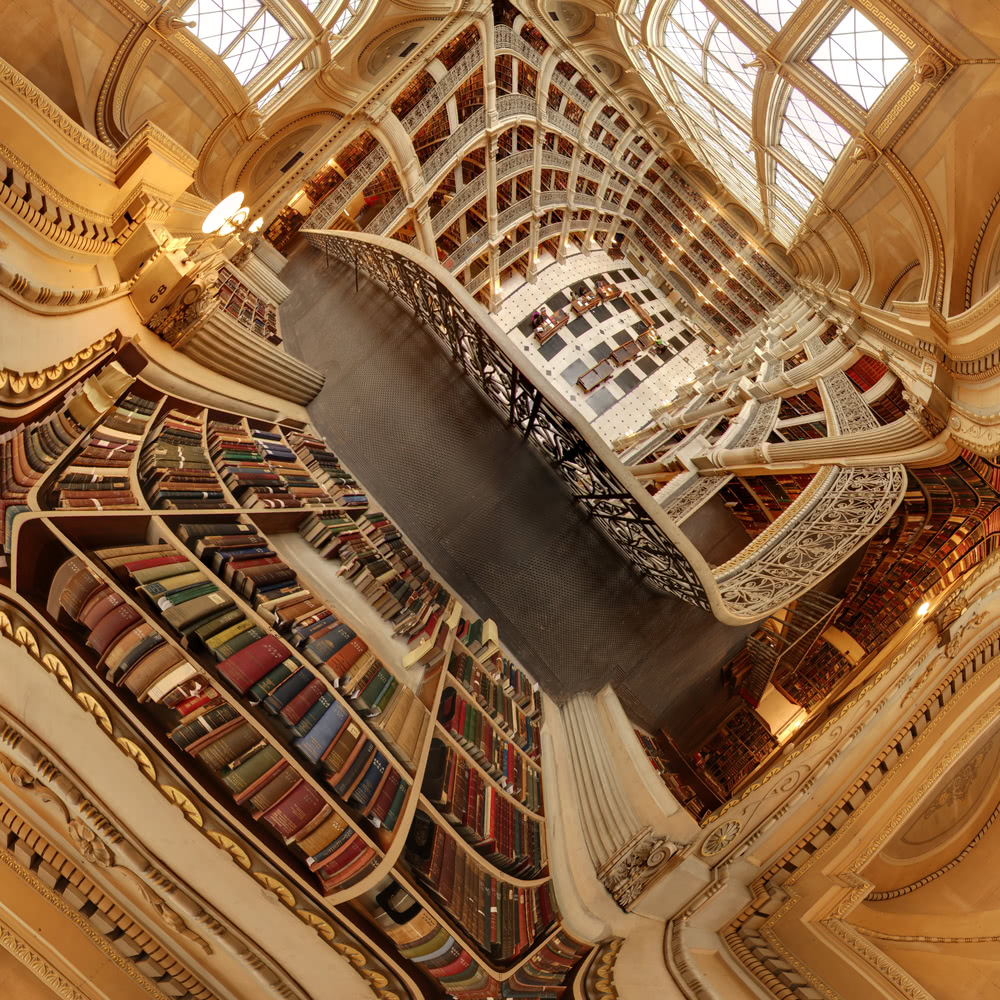}\hspace*{1em}\includegraphics[width=2.1in]{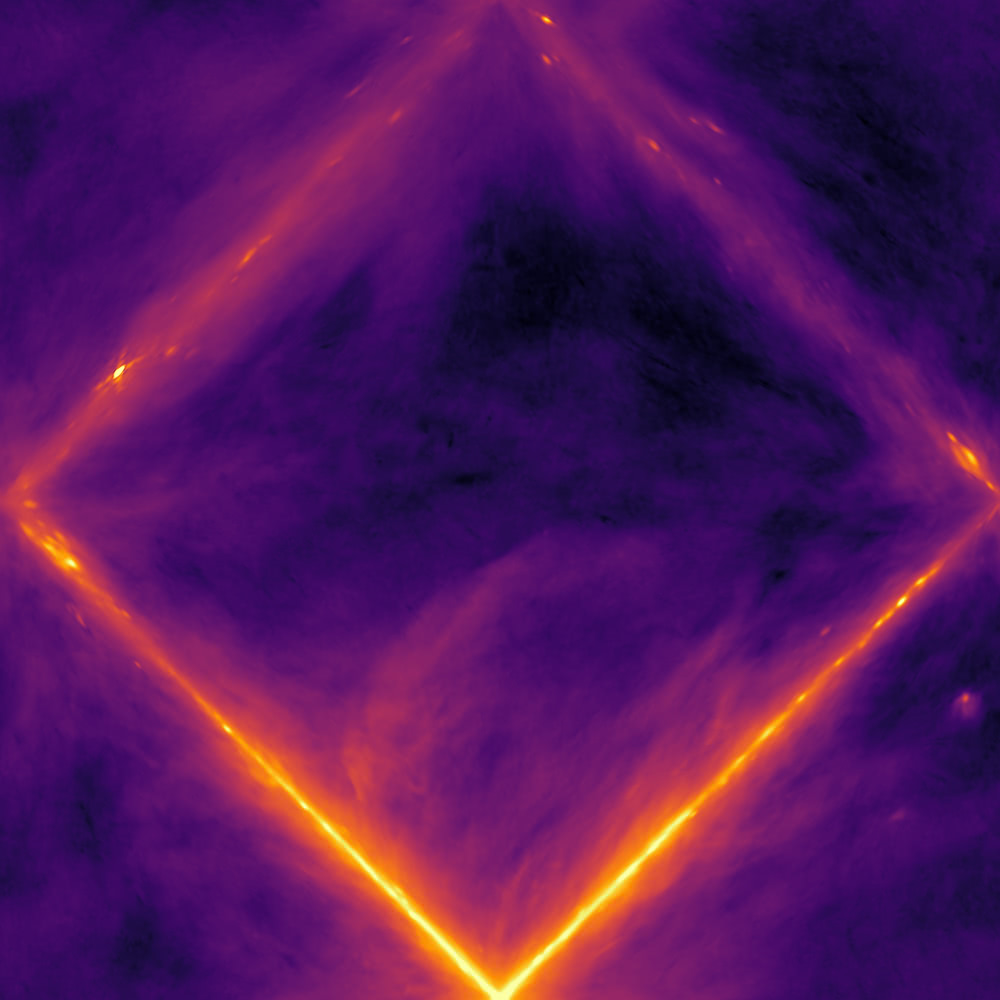}
\Description{A panoramic image of the George Peabody Library stacks and a 408\,MHz sky map are shown.}
\caption{Alternative uses. On the left, a panoramic image is projected with the new projection presented in this work, while on the right, the reprocessed Haslam 408\,MHz sky map \citep{Haslam1982, Remazeilles2015}, in Galactic coordinates, is shown using the same projection.}
\label{fig:pano}
\end{figure}

Alternative uses of the projection, outside of cartography, are demonstrated in Figure~\ref{fig:pano}. Unlike for world maps, where the outside of an approximate sphere is projected, full spherical panoramas and sky maps seek to project the view from the center of the sphere, an analogous operation. As can be seen in the figure, the projected panorama appears without extreme distortions, so the projected image is suitable for direct viewing. At the same time, the closed-form forward solution allows the projected image to be used as an input for the real-time rendering of rectilinear views, as is done by panorama viewers \citep{Petroff2019}. The combination of low angular distortion and area equivalence makes the projection better suited for panorama applications than the commonly used equirectangular format, which has both extreme area distortions and extreme angle distortions. The equal-area property avoids the storage of redundant information across multiple pixels, while the low angular distortion reduces the occurrence of sampling artifacts in rectilinear views \citep{Yan2016}. Taken together, these properties allow for reduced image file sizes, which reduce both storage and data transfer requirements for websites displaying panoramas. As current GPU texture size limitations are set with regard to square textures, a square projection maximizes the amount of image information that can be stored in a single texture; this is particularly important when displaying panoramic video, where the use of multiple textures is ideally avoided for performance reasons. The seamless tiling property can also potentially be used to avoid boundary artifacts induced by some lossy image compression codecs at the edges of an image.

Sky maps are similar to panoramas except that they show the celestial sphere instead of a view of nearby objects. Figure~\ref{fig:pano} also shows a projection of the reprocessed Haslam 408\,MHz sky map \citep{Haslam1982, Remazeilles2015}, in Galactic coordinates, as an example of this. This view of the radio sky primarily traces emission by Galactic synchrotron emission, which is brightest along the Galactic plane, in particular in the direction of the Galactic center.

\section{Conclusions}
\label{sec:conclusions}

The new equal-area quincuncial projection presented in this work combines the benefits of the Collignon quincuncial projection with that of the Gringorten projection. It has closed-form forward and inverse solutions, unlike the Gringorten projection, while maintaining relatively low angular distortion when compared to the Collignon quincuncial projection; it also lacks any significant visible cusps, which is an improvement upon both existing equal-area quincuncial projections. The fully-vectorizable closed-form solutions make the new projection suitable for use in real-time visualization applications, including hardware-accelerated GPU implementations. The generalized vertex-oriented great circle projection technique used in the projection's construction can also be used to produce other map projections, such as equal-area star projections. The example implementation code used for the analysis in this manuscript has been made available \citep{Petroff2021}.

\begin{acks}
The author acknowledges support from the \grantsponsor{NSF}{National Science Foundation Division of Astronomical Sciences}{https://nsf.gov/} under grant numbers \grantnum{NSF}{1636634} and \grantnum{NSF}{1654494} and thanks several anonymous referees whose comments helped improve this manuscript. The preparation of this work made use of JAX \citep{Frostig2018} for automatic analytic differentiation in order to calculate Tissot's indicatrix and Numba \citep{Lam2015} to speed up the calculation of the Gringorten projection. The world maps in Figures~\ref{fig:map} and~\ref{fig:maps} are based on the public domain Natural Earth\footnote{\url{https://www.naturalearthdata.com/}} small-scale land polygons. Asymptote \citep{Bowman2008}, Matplotlib \citep{Hunter2007}, Healpy \citep{Zonca2019, Gorski2005}, D3.js,\footnote{\url{https://d3js.org/}} and Inkscape\footnote{\url{https://inkscape.org/}} were used in the preparation of the included figures. The reprocessed Haslam 408\,MHz map data used in Figure~\ref{fig:pano} were downloaded from the Legacy Archive for Microwave Background Data Analysis (LAMBDA), part of the High Energy Astrophysics Science Archive Center (HEASARC); HEASARC/LAMBDA is a service of the Astrophysics Science Division at the NASA Goddard Space Flight Center.
\end{acks}

\bibliographystyle{ACM-Reference-Format}
\bibliography{paper.bib}


\begin{thebibliography}{28}


\ifx \showCODEN    \undefined \def \showCODEN     #1{\unskip}     \fi
\ifx \showDOI      \undefined \def \showDOI       #1{#1}\fi
\ifx \showISBNx    \undefined \def \showISBNx     #1{\unskip}     \fi
\ifx \showISBNxiii \undefined \def \showISBNxiii  #1{\unskip}     \fi
\ifx \showISSN     \undefined \def \showISSN      #1{\unskip}     \fi
\ifx \showLCCN     \undefined \def \showLCCN      #1{\unskip}     \fi
\ifx \shownote     \undefined \def \shownote      #1{#1}          \fi
\ifx \showarticletitle \undefined \def \showarticletitle #1{#1}   \fi
\ifx \showURL      \undefined \def \showURL       {\relax}        \fi
\providecommand\bibfield[2]{#2}
\providecommand\bibinfo[2]{#2}
\providecommand\natexlab[1]{#1}
\providecommand\showeprint[2][]{arXiv:#2}

\bibitem[\protect\citeauthoryear{Arden-Close}{Arden-Close}{1947}]%
        {Close1947}
\bibfield{author}{\bibinfo{person}{Charles Arden-Close}.}
  \bibinfo{year}{1947}\natexlab{}.
\newblock \bibinfo{booktitle}{\emph{Geographical By-ways and some other
  Geographical Essays}}.
\newblock \bibinfo{publisher}{Edward Arnold \& Co.}, \bibinfo{pages}{87--88}.
\newblock


\bibitem[\protect\citeauthoryear{Baselga}{Baselga}{2018}]%
        {Baselga2018}
\bibfield{author}{\bibinfo{person}{Sergio Baselga}.}
  \bibinfo{year}{2018}\natexlab{}.
\newblock \showarticletitle{Fibonacci lattices for the evaluation and
  optimization of map projections}.
\newblock \bibinfo{journal}{\emph{Computers {\&} Geosciences}}
  \bibinfo{volume}{117} (\bibinfo{date}{Aug.} \bibinfo{year}{2018}),
  \bibinfo{pages}{1--8}.
\newblock
\urldef\tempurl%
\url{https://doi.org/10.1016/j.cageo.2018.04.012}
\showDOI{\tempurl}


\bibitem[\protect\citeauthoryear{Bowman and Hammerlindl}{Bowman and
  Hammerlindl}{2008}]%
        {Bowman2008}
\bibfield{author}{\bibinfo{person}{John~C. Bowman} {and} \bibinfo{person}{Andy
  Hammerlindl}.} \bibinfo{year}{2008}\natexlab{}.
\newblock \showarticletitle{Asymptote: A vector graphics language}.
\newblock \bibinfo{journal}{\emph{TUGboat}} \bibinfo{volume}{29},
  \bibinfo{number}{2} (\bibinfo{year}{2008}), \bibinfo{pages}{288--294}.
\newblock
\urldef\tempurl%
\url{https://www.tug.org/TUGboat/tb29-2/tb92bowman.pdf}
\showURL{%
\tempurl}


\bibitem[\protect\citeauthoryear{Cogley}{Cogley}{2002}]%
        {Cogley2002}
\bibfield{author}{\bibinfo{person}{J.~Graham Cogley}.}
  \bibinfo{year}{2002}\natexlab{}.
\newblock \showarticletitle{Variations of the Gringorten Square Equal-area Map
  Projection}.
\newblock \bibinfo{journal}{\emph{Cartography and Geographic Information
  Science}} \bibinfo{volume}{29}, \bibinfo{number}{4} (\bibinfo{date}{Jan.}
  \bibinfo{year}{2002}), \bibinfo{pages}{381--390}.
\newblock
\urldef\tempurl%
\url{https://doi.org/10.1559/152304002782008341}
\showDOI{\tempurl}


\bibitem[\protect\citeauthoryear{Collignon}{Collignon}{1865}]%
        {Collignon1865}
\bibfield{author}{\bibinfo{person}{\'Edouard Collignon}.}
  \bibinfo{year}{1865}\natexlab{}.
\newblock \showarticletitle{Recherches sur la Repr\'esentation Plane de la
  Surface du Globe Terrestre}.
\newblock \bibinfo{journal}{\emph{Journal de l'\'Ecole Imp\'eriale
  Polytechnique}}  \bibinfo{volume}{24} (\bibinfo{year}{1865}),
  \bibinfo{pages}{73--163}.
\newblock
\urldef\tempurl%
\url{https://hdl.handle.net/2027/coo.31924069342180}
\showURL{%
\tempurl}


\bibitem[\protect\citeauthoryear{Frostig, Johnson, and Leary}{Frostig
  et~al\mbox{.}}{2018}]%
        {Frostig2018}
\bibfield{author}{\bibinfo{person}{Roy Frostig}, \bibinfo{person}{Matthew
  Johnson}, {and} \bibinfo{person}{Chris Leary}.}
  \bibinfo{year}{2018}\natexlab{}.
\newblock \showarticletitle{Compiling machine learning programs via high-level
  tracing}. In \bibinfo{booktitle}{\emph{SysML 2018}}.
\newblock
\urldef\tempurl%
\url{https://mlsys.org/Conferences/2019/doc/2018/146.pdf}
\showURL{%
\tempurl}


\bibitem[\protect\citeauthoryear{German, d'Angelo, Gross, and Postle}{German
  et~al\mbox{.}}{2007}]%
        {German2007}
\bibfield{author}{\bibinfo{person}{Daniel~M. German}, \bibinfo{person}{Pablo
  d'Angelo}, \bibinfo{person}{Michael Gross}, {and} \bibinfo{person}{Bruno
  Postle}.} \bibinfo{year}{2007}\natexlab{}.
\newblock \showarticletitle{{New Methods to Project Panoramas for Practical and
  Aesthetic Purposes}}. In \bibinfo{booktitle}{\emph{Computational Aesthetics
  in Graphics, Visualization, and Imaging}},
  \bibfield{editor}{\bibinfo{person}{Douglas~W. Cunningham},
  \bibinfo{person}{Gary Meyer}, {and} \bibinfo{person}{Laszlo Neumann}} (Eds.).
  \bibinfo{publisher}{The Eurographics Association}.
\newblock
\showISBNx{978-3-905673-43-2}
\showISSN{1816-0859}
\urldef\tempurl%
\url{https://doi.org/10.2312/COMPAESTH/COMPAESTH07/015-022}
\showDOI{\tempurl}


\bibitem[\protect\citeauthoryear{Gorski, Hivon, Banday, Wandelt, Hansen,
  Reinecke, and Bartelmann}{Gorski et~al\mbox{.}}{2005}]%
        {Gorski2005}
\bibfield{author}{\bibinfo{person}{K.~M. Gorski}, \bibinfo{person}{E. Hivon},
  \bibinfo{person}{A.~J. Banday}, \bibinfo{person}{B.~D. Wandelt},
  \bibinfo{person}{F.~K. Hansen}, \bibinfo{person}{M. Reinecke}, {and}
  \bibinfo{person}{M. Bartelmann}.} \bibinfo{year}{2005}\natexlab{}.
\newblock \showarticletitle{{HEALPix}: A Framework for High-Resolution
  Discretization and Fast Analysis of Data Distributed on the Sphere}.
\newblock \bibinfo{journal}{\emph{The Astrophysical Journal}}
  \bibinfo{volume}{622}, \bibinfo{number}{2} (\bibinfo{date}{April}
  \bibinfo{year}{2005}), \bibinfo{pages}{759--771}.
\newblock
\urldef\tempurl%
\url{https://doi.org/10.1086/427976}
\showDOI{\tempurl}
\showeprint[arxiv]{astro-ph/0409513}


\bibitem[\protect\citeauthoryear{Gringorten}{Gringorten}{1972}]%
        {Gringorten1972}
\bibfield{author}{\bibinfo{person}{Irving~I. Gringorten}.}
  \bibinfo{year}{1972}\natexlab{}.
\newblock \showarticletitle{A Square Equal-Area Map of the World}.
\newblock \bibinfo{journal}{\emph{Journal of Applied Meteorology}}
  \bibinfo{volume}{11}, \bibinfo{number}{5} (\bibinfo{date}{Aug.}
  \bibinfo{year}{1972}), \bibinfo{pages}{763--767}.
\newblock
\urldef\tempurl%
\url{https://doi.org/10.1175/1520-0450(1972)011<0763:aseamo>2.0.co;2}
\showDOI{\tempurl}


\bibitem[\protect\citeauthoryear{{Haslam}, {Salter}, {Stoffel}, and
  {Wilson}}{{Haslam} et~al\mbox{.}}{1982}]%
        {Haslam1982}
\bibfield{author}{\bibinfo{person}{C.~G.~T. {Haslam}}, \bibinfo{person}{C.~J.
  {Salter}}, \bibinfo{person}{H. {Stoffel}}, {and} \bibinfo{person}{W.~E.
  {Wilson}}.} \bibinfo{year}{1982}\natexlab{}.
\newblock \showarticletitle{{A 408 MHz all-sky continuum survey. II. The atlas
  of contour maps.}}
\newblock \bibinfo{journal}{\emph{Astronomy \& Astrophysics Supplement Series}}
   \bibinfo{volume}{47} (\bibinfo{date}{Jan.} \bibinfo{year}{1982}),
  \bibinfo{pages}{1--143}.
\newblock


\bibitem[\protect\citeauthoryear{Holho{\c{s}} and Ro{\c{s}}ca}{Holho{\c{s}} and
  Ro{\c{s}}ca}{2014}]%
        {Holhos2014}
\bibfield{author}{\bibinfo{person}{Adrian Holho{\c{s}}} {and}
  \bibinfo{person}{Daniela Ro{\c{s}}ca}.} \bibinfo{year}{2014}\natexlab{}.
\newblock \showarticletitle{An octahedral equal area partition of the sphere
  and near optimal configurations of points}.
\newblock \bibinfo{journal}{\emph{Computers {\&} Mathematics with
  Applications}} \bibinfo{volume}{67}, \bibinfo{number}{5}
  (\bibinfo{date}{March} \bibinfo{year}{2014}), \bibinfo{pages}{1092--1107}.
\newblock
\urldef\tempurl%
\url{https://doi.org/10.1016/j.camwa.2014.01.003}
\showDOI{\tempurl}


\bibitem[\protect\citeauthoryear{Huang, Shibasaki, Kasuya, and Takagi}{Huang
  et~al\mbox{.}}{1998}]%
        {Huang1998}
\bibfield{author}{\bibinfo{person}{Shaobo Huang}, \bibinfo{person}{Ryosuke
  Shibasaki}, \bibinfo{person}{Masahiro Kasuya}, {and}
  \bibinfo{person}{Masataka Takagi}.} \bibinfo{year}{1998}\natexlab{}.
\newblock \showarticletitle{Comparative study on spherical tessellation schemes
  for global {GIS}}.
\newblock \bibinfo{journal}{\emph{Geocarto International}}
  \bibinfo{volume}{13}, \bibinfo{number}{1} (\bibinfo{date}{March}
  \bibinfo{year}{1998}), \bibinfo{pages}{3--14}.
\newblock
\urldef\tempurl%
\url{https://doi.org/10.1080/10106049809354623}
\showDOI{\tempurl}


\bibitem[\protect\citeauthoryear{Hunter}{Hunter}{2007}]%
        {Hunter2007}
\bibfield{author}{\bibinfo{person}{John~D. Hunter}.}
  \bibinfo{year}{2007}\natexlab{}.
\newblock \showarticletitle{Matplotlib: A 2D Graphics Environment}.
\newblock \bibinfo{journal}{\emph{Computing in Science \& Engineering}}
  \bibinfo{volume}{9}, \bibinfo{number}{3} (\bibinfo{year}{2007}),
  \bibinfo{pages}{90--95}.
\newblock
\urldef\tempurl%
\url{https://doi.org/10.1109/mcse.2007.55}
\showDOI{\tempurl}


\bibitem[\protect\citeauthoryear{Lam, Pitrou, and Seibert}{Lam
  et~al\mbox{.}}{2015}]%
        {Lam2015}
\bibfield{author}{\bibinfo{person}{Siu~Kwan Lam}, \bibinfo{person}{Antoine
  Pitrou}, {and} \bibinfo{person}{Stanley Seibert}.}
  \bibinfo{year}{2015}\natexlab{}.
\newblock \showarticletitle{Numba: a {LLVM}-based {Python} {JIT} compiler}. In
  \bibinfo{booktitle}{\emph{LLVM '15}}. \bibinfo{publisher}{{ACM} Press}.
\newblock
\urldef\tempurl%
\url{https://doi.org/10.1145/2833157.2833162}
\showDOI{\tempurl}


\bibitem[\protect\citeauthoryear{Maurer}{Maurer}{1935}]%
        {Maurer1935}
\bibfield{author}{\bibinfo{person}{Hans Maurer}.}
  \bibinfo{year}{1935}\natexlab{}.
\newblock \bibinfo{booktitle}{\emph{Ebene Kugelbilder: Ein Linn\'esches System
  der Kartenentw\"urfe}}.
\newblock Number 221 in \bibinfo{series}{Petermanns Mitteilungen}.
  \bibinfo{publisher}{Justus Perthes}.
\newblock


\bibitem[\protect\citeauthoryear{McGlynn, Fay, Wong, and Rosenfield}{McGlynn
  et~al\mbox{.}}{2019}]%
        {McGlynn2019}
\bibfield{author}{\bibinfo{person}{Thomas McGlynn}, \bibinfo{person}{Jonathan
  Fay}, \bibinfo{person}{Curtis Wong}, {and} \bibinfo{person}{Philip
  Rosenfield}.} \bibinfo{year}{2019}\natexlab{}.
\newblock \showarticletitle{Octahedron-based Projections as Intermediate
  Representations for Computer Imaging: {TOAST}, {TEA}, and More}.
\newblock \bibinfo{journal}{\emph{The Astrophysical Journal Supplement Series}}
  \bibinfo{volume}{240}, \bibinfo{number}{2} (\bibinfo{date}{Jan.}
  \bibinfo{year}{2019}), \bibinfo{pages}{22}.
\newblock
\urldef\tempurl%
\url{https://doi.org/10.3847/1538-4365/aaf79e}
\showDOI{\tempurl}
\showeprint[arxiv]{1812.03926}~[astro-ph.IM]


\bibitem[\protect\citeauthoryear{Petroff}{Petroff}{2019}]%
        {Petroff2019}
\bibfield{author}{\bibinfo{person}{Matthew~A. Petroff}.}
  \bibinfo{year}{2019}\natexlab{}.
\newblock \showarticletitle{Pannellum: a lightweight web-based panorama
  viewer}.
\newblock \bibinfo{journal}{\emph{Journal of Open Source Software}}
  \bibinfo{volume}{4}, \bibinfo{number}{40} (\bibinfo{date}{Aug.}
  \bibinfo{year}{2019}), \bibinfo{pages}{1628}.
\newblock
\urldef\tempurl%
\url{https://doi.org/10.21105/joss.01628}
\showDOI{\tempurl}


\bibitem[\protect\citeauthoryear{Petroff}{Petroff}{2021}]%
        {Petroff2021}
\bibfield{author}{\bibinfo{person}{Matthew~A. Petroff}.}
  \bibinfo{year}{2021}\natexlab{}.
\newblock \bibinfo{title}{Supplement to \emph{A Square Equal-area Map
  Projection with Low Angular Distortion, Minimal Cusps, and Closed-form
  Solutions}}.
\newblock
\newblock
\urldef\tempurl%
\url{https://doi.org/10.5281/zenodo.4661351}
\showDOI{\tempurl}


\bibitem[\protect\citeauthoryear{{Remazeilles}, {Dickinson}, {Banday},
  {Bigot-Sazy}, and {Ghosh}}{{Remazeilles} et~al\mbox{.}}{2015}]%
        {Remazeilles2015}
\bibfield{author}{\bibinfo{person}{M. {Remazeilles}}, \bibinfo{person}{C.
  {Dickinson}}, \bibinfo{person}{A.~J. {Banday}}, \bibinfo{person}{M.~A.
  {Bigot-Sazy}}, {and} \bibinfo{person}{T. {Ghosh}}.}
  \bibinfo{year}{2015}\natexlab{}.
\newblock \showarticletitle{{An improved source-subtracted and destriped
  408-{MHz} all-sky map}}.
\newblock \bibinfo{journal}{\emph{Monthly Notices of the Royal Astronomical
  Society}} \bibinfo{volume}{451}, \bibinfo{number}{4} (\bibinfo{date}{Aug.}
  \bibinfo{year}{2015}), \bibinfo{pages}{4311--4327}.
\newblock
\urldef\tempurl%
\url{https://doi.org/10.1093/mnras/stv1274}
\showDOI{\tempurl}
\showeprint[arxiv]{1411.3628}~[astro-ph.IM]


\bibitem[\protect\citeauthoryear{Ro{\c{s}}ca}{Ro{\c{s}}ca}{2011}]%
        {Rosca2011}
\bibfield{author}{\bibinfo{person}{Daniela Ro{\c{s}}ca}.}
  \bibinfo{year}{2011}\natexlab{}.
\newblock \showarticletitle{Uniform and refinable grids on elliptic domains and
  on some surfaces of revolution}.
\newblock \bibinfo{journal}{\emph{Appl. Math. Comput.}} \bibinfo{volume}{217},
  \bibinfo{number}{19} (\bibinfo{date}{June} \bibinfo{year}{2011}),
  \bibinfo{pages}{7812--7817}.
\newblock
\urldef\tempurl%
\url{https://doi.org/10.1016/j.amc.2011.02.095}
\showDOI{\tempurl}


\bibitem[\protect\citeauthoryear{Snyder}{Snyder}{1987}]%
        {Snyder1987}
\bibfield{author}{\bibinfo{person}{John~P. Snyder}.}
  \bibinfo{year}{1987}\natexlab{}.
\newblock \bibinfo{title}{Map projections: A working manual}.
\newblock
\newblock
\urldef\tempurl%
\url{https://doi.org/10.3133/pp1395}
\showDOI{\tempurl}


\bibitem[\protect\citeauthoryear{Snyder}{Snyder}{1993}]%
        {Snyder1993}
\bibfield{author}{\bibinfo{person}{John~P. Snyder}.}
  \bibinfo{year}{1993}\natexlab{}.
\newblock \bibinfo{booktitle}{\emph{Flattening the Earth: Two Thousand Years of
  Map Projections}}.
\newblock \bibinfo{publisher}{University of Chicago Press},
  \bibinfo{pages}{114}.
\newblock
\showISBNx{0-226-76746-9}


\bibitem[\protect\citeauthoryear{Tissot}{Tissot}{1881}]%
        {Tissot1881}
\bibfield{author}{\bibinfo{person}{A. Tissot}.}
  \bibinfo{year}{1881}\natexlab{}.
\newblock \bibinfo{booktitle}{\emph{M\'emoire sur la Repr\'esentation des
  Surfaces et les Projections des Cartes G\'eographiques}}.
\newblock \bibinfo{publisher}{Gauthier-Villars}.
\newblock


\bibitem[\protect\citeauthoryear{Tobler and Chen}{Tobler and Chen}{1986}]%
        {Tobler1986}
\bibfield{author}{\bibinfo{person}{Waldo Tobler} {and}
  \bibinfo{person}{{Zi-tan} Chen}.} \bibinfo{year}{1986}\natexlab{}.
\newblock \showarticletitle{A Quadtree for Global Information Storage}.
\newblock \bibinfo{journal}{\emph{Geographical Analysis}} \bibinfo{volume}{18},
  \bibinfo{number}{4} (\bibinfo{date}{Oct.} \bibinfo{year}{1986}),
  \bibinfo{pages}{360--371}.
\newblock
\urldef\tempurl%
\url{https://doi.org/10.1111/j.1538-4632.1986.tb00108.x}
\showDOI{\tempurl}


\bibitem[\protect\citeauthoryear{van Leeuwen and Strebe}{van Leeuwen and
  Strebe}{2006}]%
        {vanLeeuwen2006}
\bibfield{author}{\bibinfo{person}{Diederik van Leeuwen} {and}
  \bibinfo{person}{Daniel Strebe}.} \bibinfo{year}{2006}\natexlab{}.
\newblock \showarticletitle{A ``Slice-and-Dice'' Approach to Area Equivalence
  in Polyhedral Map Projections}.
\newblock \bibinfo{journal}{\emph{Cartography and Geographic Information
  Science}} \bibinfo{volume}{33}, \bibinfo{number}{4} (\bibinfo{date}{Jan.}
  \bibinfo{year}{2006}), \bibinfo{pages}{269--286}.
\newblock
\urldef\tempurl%
\url{https://doi.org/10.1559/152304006779500687}
\showDOI{\tempurl}


\bibitem[\protect\citeauthoryear{Warntz}{Warntz}{1968}]%
        {Warntz1968}
\bibfield{editor}{\bibinfo{person}{William Warntz}} (Ed.).
  \bibinfo{year}{1968}\natexlab{}.
\newblock \bibinfo{booktitle}{\emph{Plane Globe Projection: a Linnean System of
  Map Projection}}.
\newblock Number~23 in \bibinfo{series}{Harvard Papers in Theoretical
  Geography: Geography and the Properties of Surfaces}.
  \bibinfo{publisher}{Laboratory for Computer Graphics and Spatial Analysis,
  Harvard}, \bibinfo{pages}{166}.
\newblock
\urldef\tempurl%
\url{https://apps.dtic.mil/sti/citations/AD0675812}
\showURL{%
\tempurl}
\newblock
\shownote{English translation of Maurer (1935).}


\bibitem[\protect\citeauthoryear{Yan, Song, and Gong}{Yan
  et~al\mbox{.}}{2016}]%
        {Yan2016}
\bibfield{author}{\bibinfo{person}{Jin Yan}, \bibinfo{person}{Xiao Song}, {and}
  \bibinfo{person}{Guanghong Gong}.} \bibinfo{year}{2016}\natexlab{}.
\newblock \showarticletitle{Averaged ratio between complementary profiles for
  evaluating shape distortions of map projections and spherical hierarchical
  tessellations}.
\newblock \bibinfo{journal}{\emph{Computers {\&} Geosciences}}
  \bibinfo{volume}{87} (\bibinfo{date}{Feb.} \bibinfo{year}{2016}),
  \bibinfo{pages}{41--55}.
\newblock
\urldef\tempurl%
\url{https://doi.org/10.1016/j.cageo.2015.11.009}
\showDOI{\tempurl}


\bibitem[\protect\citeauthoryear{Zonca, Singer, Lenz, Reinecke, Rosset, Hivon,
  and Gorski}{Zonca et~al\mbox{.}}{2019}]%
        {Zonca2019}
\bibfield{author}{\bibinfo{person}{Andrea Zonca}, \bibinfo{person}{Leo Singer},
  \bibinfo{person}{Daniel Lenz}, \bibinfo{person}{Martin Reinecke},
  \bibinfo{person}{Cyrille Rosset}, \bibinfo{person}{Eric Hivon}, {and}
  \bibinfo{person}{Krzysztof Gorski}.} \bibinfo{year}{2019}\natexlab{}.
\newblock \showarticletitle{Healpy: equal area pixelization and spherical
  harmonics transforms for data on the sphere in Python}.
\newblock \bibinfo{journal}{\emph{Journal of Open Source Software}}
  \bibinfo{volume}{4}, \bibinfo{number}{35} (\bibinfo{date}{March}
  \bibinfo{year}{2019}), \bibinfo{pages}{1298}.
\newblock
\urldef\tempurl%
\url{https://doi.org/10.21105/joss.01298}
\showDOI{\tempurl}


\end{thebibliography}

\appendix
\section{Inverse Mapping}
\label{app:inverse}

Here, the inverse mapping is derived, using the previously defined $\phi_0$. Although the forward mapping equations are not directly invertible, an inverse mapping can be derived by performing the ``slice-and-dice'' technique in reverse, by matching area ratios on the spherical sub-triangles to those on the Euclidean sub-triangles. Starting with projected map coordinates $x_m,y_m \in [-1, 1]$,

\begin{equation}
q = \begin{cases}
0 & x_m \geq 0, y_m \leq 0 \\
1 & x_m > 0, y_m > 0 \\
2 & x_m \leq 0, y_m > 0 \\
3 & x_m < 0, y_m \leq 0
\end{cases}
\end{equation}
\begin{equation}
\zeta = \frac{\pi}{4} + \frac{\pi}{2}q
\end{equation}
\begin{equation}
\begin{pmatrix}
x_c \\ y_c
\end{pmatrix} = \begin{pmatrix}
\sqrt{6}(x_m\cos\zeta + y_m\sin\zeta) \\
3\sqrt{2}(y_m\cos\zeta - x_m\sin\zeta)
\end{pmatrix}
\end{equation}
gives Cartesian coordinates relative to the Euclidean equilateral triangle face center. We then separate the northern and southern hemispheres with
\begin{equation}
\phi_\mathrm{sgn} = \begin{cases}
1 & y_c \geq -3 \\
-1 & yc < -3
\end{cases}
\end{equation}
and define a hemisphere-independent coordinate with
\begin{equation}
y_h = \begin{cases}
y_c & y_c \geq -3 \\
-6 - y_c & y_c < -3.
\end{cases}
\end{equation}
The corresponding polar coordinates are then
\begin{align}
r' &= \sqrt{x_c^2 + (h'-3 - y_h)^2} \\
\theta' &= \left|\arctantwo(x_c, h'-3 - y_h)\right|,
\end{align}
where $h'$ is defined by Equation~(\ref{eq:hprime}). Using $\psi_0'$, $\psi_1'$, and $\psi_2'$ defined by Equations (\ref{eq:psi0prime}), (\ref{eq:psi1prime}), (\ref{eq:psi2prime}), respectively, we find
\begin{equation}
\alpha' = \begin{cases}
\theta' & \theta' \leq \psi_0' \\
\pi - \psi_2' - \theta' & \psi_0' < \theta' \leq \psi_0' + \psi_1' \\
\theta' + \psi_2' - \pi & \psi_0' + \psi_1' < \theta'.
\end{cases}
\end{equation}
The fixed sub-triangle lengths $c$, $G$, $G'$, $F$, $a'$, and $c'$ are defined by Equations (\ref{eq:c}), (\ref{eq:G}), (\ref{eq:Gprime}), (\ref{eq:F}), (\ref{eq:aprime}), and (\ref{eq:cprime}), respectively, except with $\theta$ replaced with $\theta'$, $\psi_0$ replaced with $\psi_0'$, and $\psi_1$ replaced with $\psi_1'$, in the case-condition statements only. The remaining necessary fixed sub-triangle length is defined with
\begin{equation}
b = \begin{cases}
\pi / 4 & \theta' \leq \psi_0' \\
\arctan\left(\sqrt 2\tan\phi_0\right) & \psi_0' < \theta' \leq \psi_0' + \psi_1' \\
\pi / 2 - \arctan\left(\sqrt 2\tan\phi_0\right) & \psi_0' + \psi_1' < \theta'.
\end{cases}
\end{equation}
Other angles and lengths of the Euclidean sub-triangles can then be found with
\begin{equation}
\beta' = G' - \alpha'
\end{equation}
\begin{equation}
x' = \sqrt{r'^2 + c'^2 - 2r'c'\cos\beta'}
\end{equation}
\begin{equation}
\gamma' = \arccos\left(\frac{r'^2-x'^2-c'^2}{-2x'c'}\right)
\end{equation}
\begin{equation}
\epsilon' = \pi - G' - \gamma'
\end{equation}
\begin{equation}
y' = \frac{r'\sin\alpha'}{\sin\epsilon'}
\end{equation}
\begin{equation}
u' = \sqrt{c'^2 + (x' + y')^2-2c'(x'+y')\cos\gamma'}
\end{equation}
\begin{equation}
v' = a' - u'.
\end{equation}
Next, the ``slice'' operation is performed in reverse to match the area ratio of the spherical sub-triangles to that of the Euclidean sub-triangles, with
\begin{equation}
\frac{v'}{a'} = \frac{\delta + \arccos(\sin\delta\cos b) - \pi / 2}{F + G - \pi / 2}
\end{equation}
yielding
\begin{equation}
\delta = \arctan\left(\frac{-\sin[v'(F+G-\pi/2)/a']}{\cos b - \cos[v'(F+G-\pi/2)/a']}\right)
\end{equation}
when solved for $\delta$. Other parameters of the spherical sub-triangles are then found with
\begin{equation}
\gamma = F - \delta
\end{equation}
\begin{equation}
\begin{split}
\cos(x+y) &= \cos\left[\arctan\left(\frac{\tan b}{\cos\delta}\right)\right] \\
&= 1 \left/ \sqrt{1 + \left(\frac{\tan b}{\cos\delta}\right)^2}\right..
\end{split}
\end{equation}
Next, the ``dice'' step is reversed, with Equation~(\ref{eq:dice}) yielding
\begin{equation}
x = \arccos\left[1 - \left(\frac{x'}{(x'+y')}\right)^2[1-\cos(x+y)]\right]
\end{equation}
when solved for $x$. A distance from and an azimuth around the spherical triangle center point are then found with
\begin{equation}
r = \arccos(\cos x\cos c + \sin x \sin c \cos\gamma)
\end{equation}
\begin{equation}
\beta = \arcsin\left(\frac{\sin x\sin\gamma}{\sin r}\right)
\end{equation}
\begin{equation}
\alpha = \begin{cases}
\psi_0 - \beta & \theta' \leq \psi_0' \\
\beta + \psi_0 & \psi_0' < \theta' \leq \psi_0' + \psi_1' \\
\pi - \beta & \psi_0' + \psi_1' < \theta'.
\end{cases}
\end{equation}
Finally, these are converted to latitude, $\phi$, and longitude, $\lambda$, with
\begin{equation}
\phi_h = \arcsin(\sin\phi_0\cos r - \cos\phi_0\sin r\cos\alpha)
\end{equation}
\begin{equation}
\lambda_0 = \frac{\pi}{4} + \frac{\pi}{2}q
\end{equation}
\begin{equation}
\phi = \phi_\mathrm{sgn}\phi_h
\end{equation}
\begin{equation}
\lambda = \lambda_0 + \sgn y_c \arctan\left(\frac{\sin\alpha\sin r\cos\phi_0}{\cos r - \sin\phi_0\sin\phi_h}\right).
\end{equation}

\section{Collignon quincuncial projection}
\label{app:collignon}

The Collignon quincuncial projection forward mapping equations used in this article's comparisons are presented below. Starting with latitude, $\phi \in [-\pi/2, \pi/2]$, and longitude, $\lambda \in [0, 2\pi)$ yields map Cartesian coordinates $x_t, y_t \in [-1, 1]$:

\begin{equation}
q = \left\lfloor \frac{2}{\pi}\lambda \right\rfloor
\end{equation}
\begin{equation}
\lambda_0 = \lambda - \frac{\pi}{2}q - \frac{\pi}{4}
\end{equation}
\begin{equation}
\cos\phi = \cos\left(\frac{|\phi|}{2} + \frac{\pi}{4}\right)
\end{equation}
\begin{equation}
x = -\frac{2\sqrt{2}}{\pi}\lambda_0 \cos\phi
\end{equation}
\begin{equation}
y = \begin{cases}
1 -\cos\phi / \sqrt{2} & \phi < 0 \\
\cos\phi / \sqrt{2} & \phi \geq 0
\end{cases}
\end{equation}
\begin{equation}
x_t = \begin{cases}
\hphantom{-}x - y & q = 0 \\
-x - y & q = 1 \\
-x + y & q = 2 \\
\hphantom{-}x + y & q = 3
\end{cases}
\end{equation}
\begin{equation}
y_t = \begin{cases}
\hphantom{-}x + y & q = 0 \\
\hphantom{-}x - y & q = 1 \\
-x - y & q = 2 \\
-x + y & q = 3.
\end{cases}
\end{equation}
The map coordinate system matches that of the new projection presented in Section~\ref{sec:projection}, including the area scale factor $1/\pi$, but variables used in this Appendix should be considered separate from the rest of the article.

\end{document}